# Comparative Study of APIs and Frameworks for Haptic Application Development

Dorin M. Popovici, Felix G. Hamza-Lup, Adrian Seitan, Crenguta M. Bogdan
Mathematics and Computer Science Department
Ovidius University
Constanta, Romania
{dmpopovici, cbogdan}@univ-ovidius.ro, felix.hamza-lup@armstrong.edu, seitan.adi@gmail.com

*Abstract*—The simulation of tactile sensation using haptic devices is increasingly investigated in conjunction with simulation and training. In this paper we explore the most popular haptic frameworks and APIs. We provide a comprehensive review and comparison of their features and capabilities, from the perspective of the need to develop a haptic simulator for medical training purposes. In order to compare the studied frameworks and APIs, we identified and applied a set of 11 criteria and we obtained a classification of platforms, from the perspective of our project. According to this classification, we used the best platform to develop a visuo-haptic prototype for liver diagnostics.

## I. INTRODUCTION

Virtual and augmented reality technologies are essential for current and future working environments with applications in a wide range of human activities, such as: medicine, engineering, education, tourism and more. In the last decade, the technology has evolved in such a way that it can offer advanced multi-modal simulations by combining visual with tactile feedback, and by doing so, it is augmenting the user's sense of presence inside the virtual environments on multiple channels (i.e. tactile, vision, auditory). The development of such systems represents a priority for international research programs due to their capability to improve the efficiency of different human activities (e.g. training activities of medical personnel, realistic simulations for testing residents in medical procedures, etc).

If we describe any human activity domain without referencing the 3$^{rd}$ millennium technology, we risk offering an incomplete image of the subject. Particularly in educational and training environments which are strongly influenced by the evolution of technology.

Sensorial renderings that are presented to a user of such an environment are multimodal, spanning from the audio sense to visual and recently to the tactile sense.

## II. THE HAPTICMED PROJECT

The main goal of the Haptic Interfaces in Medical Applications (HapticMed) project is to strengthen our team's competency in the domain of haptic interfaces. This goal will be achieved by transferring the scientific and technical expertise from the second author, expert in the design and development of user interfaces with haptic-feedback (i.e. tactile retroaction) and their usage in medical training and patient rehabilitation, to the Ovidius University team. Specifically, the project is oriented towards simulation applications in the medical field, especially in laparoscopic surgeon training [1].

The paper is structured as follows. Section 3 presents the conceptual architecture of a haptic platform, followed by a brief description of existing haptic frameworks and APIs for Haptic User Interface (HUI) development, in Section 4. In Section 5 we apply a set of 11 criteria in order to obtain a classification of these frameworks and APIs, from the perspective of our project. We conclude with discussions.

## III. HAPTIC PLATFORMS ARCHITECTURE

Each of the studied platform implements an architecture similar to the one presented in Fig. 1. It is easy to observe the central role that haptic and visual devices play in the multi-sensorial application development. Moreover, there is a need of other devices as well, that assures audio rendering for example, and may respond to other specific application needs. Each API is responsible for the implementation of the interfaces with those devices and for their synchronization with the visual component.

Each of the studied frameworks uses a different meta-language, such as XML, VRML or X3D [2] in order to describe the scene graph structure. Frequently, Python is used for enriching the scene graph structure specified by a meta-language, adding functionality by scripting modules.

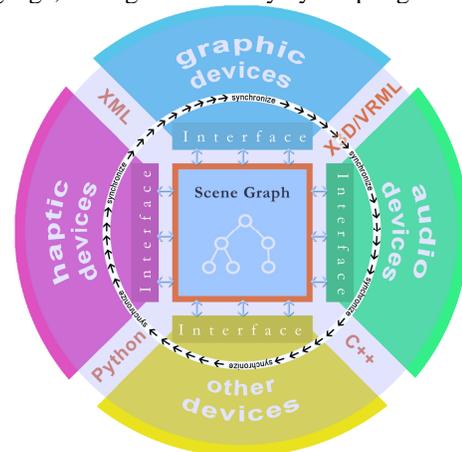

**Figure 1. Conceptual architecture of a multisensorial haptic platform**

## IV. Haptic Frameworks and APIs

Without being exhaustive in what it concerns haptic devices, our contribution attempts to be comprehensive in what it concerns existing haptic APIs.

ReachIn [3] is a programming interface, written in C++ and based on the scene graph implemented by VRML. The API was one of the first commercial ones that involve haptics. Its platform structure allows the development of multimodal interfaces and synchronizes haptic, graphic, audio or non-haptic devices.

SOFA (Simulation Open-Framework Architecture) [4] is an open-source simulation framework dedicated to the development of algorithms for deformation. Using a scene graph structure, SOFA provides several views in modeling 3D objects: a dynamic view that include masses and constitutive laws for the objects, a collision view that use simplified 3D models of the objects in collision computation, and a visual view that uses a complex 3D graphical representation. SOFA assures the scene consistency between these models by using mapping modules. Moreover, SOFA implements complex real-time algorithms that use multiple representations of the simulated objects in the three views.

Computer Haptics and Active Interfaces - CHAI3D [5] is an open-source designed to facilitate the development of 3D modeling applications augmented with haptic rendering. It supports several commercial haptic interfaces such as Servo2Go and Sensoray 626 I/O board, IEEE1394 interface. CHAI3D provides an easy solution to interface any haptic device with a specific computer-based application. CHAI3D framework allows extensions using modules for ODE [6] and dynamic engines that simulate rigid and deformable objects in real-time. Moreover CHAI3D enables the development of new classes, in order to integrate new haptic and visual rendering algorithms as well as drivers for new devices.

A popular open-source platform, H3D [7] is dedicated to haptic modeling that combines the OpenGL and X3D standards together with haptic rendering in a single scene graph that mixes haptic and graphic components. H3D is independent of haptic device multi-platform that allows audio and 3D stereoscopic device integration. H3D is conceived to support rapid prototyping. Combining X3D, C++ and the Python scripting language, H3D improves the speed of execution, when performance is critical, as well as high speed of development, when rapid prototyping is required..

General Physical Simulation Interface (GiPSi) [8] is an open-source framework that presents a flexible architecture, developed to simulate surgical procedures at organ level. The architecture interconnects computational and data models, developed by different research teams, quantitative validation of biological simulations together with software modules interconnections.

OpenHaptics toolkit [9] developed by SenseAble, includes the QuickHaptics interface, the haptic device (HD) interface (HDAPI), the haptic library (HL) interface (HLAPI), together with tools and drivers for PHANTOM® devices (PDD). The toolkit is accompanied by a solid documentation and a programmer's guide.

The HDAPI provides low level access to the haptic devices. The programmer can replay forces on the device and has access to the device driver configuration settings and debugging support. The HLAPI covers haptic feedback at a higher level and requires OpenGL development knowledge. QuickHaptics allows haptic application development or extensions for existing applications.

Through HLAPI and HDAPI interfaces, OpenHaptics gives the possibility of both high-level and low-level haptic programming, by means of an adaptable control module.

In the following section, we provide a comparison among these APIs based on a set of criteria as input in a house of quality analysis [10].

## V. Comparative Study of Haptic Platforms

In the following, we will compare the presented haptic APIs in order to qualitatively evaluate them from the perspective of the HapticMed project. To this end, we adopt the House of Quality metodology [10] on the basis of 11 criteria in accordance with the mentioned project requirements. The selected criteria are:
- license type
- required resources
- multimodal resources
- compatibility with haptic devices
- 3D navigation metaphors and devices
- implementation language
- extensibility and adaptability
- real/virtual time execution
- dynamic configuration of the scene
- documentation
- availability of the API

All these criteria are evaluated for each of the studied framework using a four-scale graded value. The score is to be interpreted as follows:
- 0 means that the framework has low quality capabilities or *non-existent* capabilities for the assessed subdomain;
- 1 means that the framework meets the analyzed capability, and it is *weakly* fulfilled;
- 3 means that the framework meets the analyzed capability, and it is fulfilled at a *medium* rate;
- 9 means that the framework meets the analyzed capability, and it is *strongly* fulfilled.

This rating system was used in order to fill in the correspondent tables to each criterion. The scores were given by the programmers that have developed applications using each of the labeled frameworks.

### A. License type

The first criterion refers to the influence that the type of the license has over the developing process. The advantages of a paid license consist of the support that the framework developers can offer, but on the other hand, this kind of license adds costs that may not be eligible to the project, especially at the prototype level.

SOFA is an open-source framework written in the C++ programming language. It is a simulation framework which offers tools for implementing algorithms and also for splitting complex objects into functional components.

CHAI3D offers two types of licenses: an open license and a professional license (Professional Edition License). Open-source license is a GNU General Public License ("GPL") version 2 type. This license offers the users free access to the source-code and allows the distribution, use and change of the source-code while observing the stipulated terms in GNU GPL. If one uses CHAI3D code inside an application that is not under a GNU GPL type license, in order to sell the product the professional license must be purchased.

Another framework that is available under two types of licenses is OpenHaptics; it has an academic license and a commercial one. The academic license is used for educational or research purposes as long as the developed applications are not commercialized.

H3D is also an open-source framework available under the GNU GPL license; it allows modification and distribution of the code as long as the process is under the license terms.

Also, there are applications labeled as open-source, but they require one or more applications that work only under paid license; these are semi-open source. For example, GiPSi is an open-source framework dedicated to surgical simulations at organ level. It supports developing reusable models, adaptation of heterogeneous computing models and it offers a working area to use multiple heterogeneous models. GiPSi is independent of the models it uses and, therefore, it offers easy integration of haptic models and of the underlying processes for haptic simulations.

A commercial license framework is ReachIn. It can be used only by contacting the developing team. Official website [3] contains the characteristics of the ReachIn products and contact information.

The license oriented evaluation of the haptic frameworks is illustrated in the Table I.

### B. Ressources needed

The conditions to use a visuo-haptic framework depend on certain minimum hardware requirements, and rely on some software packages.

*1) Hardware resources*

The hardware requirements refer to the connection between the haptic device and the computer. This could be a FireWire (IEEE 1394) on 6 or 4 pins connection, USB, PCI or through a parallel port. The main developer of haptic devices, Sensable recommends the use of the IEEE 1394 connection cards with VIA chipset because of its high performance.

Minimum requirements assure the normal execution of a simple virtual scene, meaning a scene that does not contain any complex object and which does not has to process a large amount of information. More complex scenes require high performance hardware components. Therefore, the second part of the hardware requirements refers to hardware system performance. For the studied platforms, the requirements for RAM and processor speed depend on how complex the application is, while for the permanent memory, the minimum requirement consists of the memory occupied by the framework's files.

TABLE I. TYPES OF LICENSES UNDER WHICH ARE AVAILABLE THE STUDIED FRAMEWORKS

| | | | ReachIn | SOFA | CHAI3D | H3D | GiPSi | Open Haptics |
|---|---|---|---|---|---|---|---|---|
| Paid license | Open-source | View source-code | 0 | 0 | 9 | 9 | 0 | 0 |
| | | Changing existing code | 0 | 0 | 9 | 9 | 0 | 0 |
| | | Software distribution modified/ unmodified | 0 | 0 | 9 | 9 | 0 | 0 |
| | | Class extend | 0 | 0 | 9 | 9 | 0 | 0 |
| | | Community support | 0 | 0 | 1 | 9 | 0 | 0 |
| | Semi-open source | View source-code | 0 | 0 | 0 | 0 | 0 | 0 |
| | | Changing existing code | 0 | 0 | 0 | 0 | 0 | 0 |
| | | Software distribution modified/ unmodified | 0 | 0 | 0 | 0 | 0 | 0 |
| | | Class extend | 0 | 0 | 0 | 0 | 0 | 0 |
| | | Community support | 0 | 0 | 0 | 0 | 0 | 0 |
| | Closed source | View source-code | 1 | 0 | 0 | 0 | 0 | 0 |
| | | Changing existing code | 1 | 0 | 0 | 0 | 0 | 0 |
| | | Software distribution modified / unmodified | 0 | 0 | 0 | 0 | 0 | 0 |
| | | Class extend | 3 | 0 | 0 | 0 | 0 | 0 |
| Free license | Open-source | View source-code | 0 | 9 | 9 | 9 | 0 | 0 |
| | | Changing existing code | 0 | 9 | 9 | 9 | 0 | 0 |
| | | Software distribution modified / unmodified | 0 | 9 | 9 | 9 | 0 | 0 |
| | | Class extend | 0 | 9 | 9 | 9 | 0 | 0 |
| | | Community support | 0 | 1 | 1 | 9 | 0 | 0 |
| | Semi-open source | View source-code | 0 | 0 | 0 | 0 | 9 | 0 |
| | | Changing existing code | 0 | 0 | 0 | 0 | 9 | 0 |
| | | Software distribution modified / unmodified | 0 | 0 | 0 | 0 | 9 | 0 |
| | | Class extend | 0 | 0 | 0 | 0 | 9 | 0 |
| | | Community support | 0 | 0 | 0 | 0 | 1 | 0 |
| | Closed-source | View source-code | 0 | 0 | 0 | 0 | 0 | 3 |
| | | Changing existing code | 0 | 0 | 0 | 0 | 0 | 1 |
| | | Software distribution modified / unmodified | 0 | 0 | 0 | 0 | 0 | 9 |
| | | Class extend | 0 | 0 | 0 | 0 | 0 | 9 |

*2) Software resources*

The second evaluation criterion of a visuo-haptic framework is represented by the software packages required for development.

The essential software interface for connection between the haptic device and the operating system is represented by the haptic device driver. It is available online following the links to haptic device manufacturer's websites.

*a) Programming environment*

To run on the Windows OS, OpenHaptics requires Microsoft Visual Studio, version 2003 or 2005.

For the ReachIn API, the software requirements are considered to be low. If the virtual scene is built using the C++ programming language, a programming environment such as Microsoft Visual Studio is required; a version released in 2003 or 2005; if the scenes are built using the VRML standard choosing the programming environment is the developer's choice.

SOFA framework supports two developing methods: at a lower level, using C++ programming language through Microsoft Visual Studio, or at higher level, using the available graphical user interface, independent of any programming language.

To use the H3D framework, the source-code and the CMake application are required. Users can run their virtual scenes (that are built using only implemented nodes) using the H3D Viewer. For Windows OS, H3D developers have provided a compact installer facilitating the installation process.

CHAI3D application under Windows OS requires Microsoft Visual C++, any of the versions released in: 2003, 2005, 2008 or 2010 (if there is no haptic device connected, the Chai3D application will instantiate a virtual device which allows navigation and interaction in the virtual environment).

Compiling GiPSi's source code requires several libraries such as: Intel Math Kernel Library, Posix Threads, Xerces-C++ XML Parser and ACE\TAO.

All of the studied frameworks work under Linux OS, using GCC/G++ compiler (Table II).

*b) Graphic rendering library*

The frameworks we studied employ OpenGL for graphics rendering. This library is independent of the operating system (Windows / Linux / MacOS ).

For the OpenHaptics framework, besides OpenGL, DirectX can be used. The main disadvantage of the DirectX graphic library is that it is not portable. As a conclusion, all studied frameworks use the OpenGL graphic library. OpenHaptics and SOFA may be rendered using Direct3D/DirectX graphic library.

In order to create visual representations of the geometric models inside simulations, SOFA supports two methods: either through directly accessing OpenGL [11], or through using the OGRE rendering engine - Open Source 3D Graphics Engine [12]. OpenGL represents the lowest level in graphic rendering architecture, while OGRE is an object oriented architecture which offers an intuitive graphic interface when developing applications. This interface may contain in its implementation the OpenGL or Direct3D libraries provided by Microsoft. In Table II we present the grades of the studied APIs and frameworks regarding the libraries needed for simulators developing.

*C. Multimodal ressources*

This criterion is essential in the developing process using a specific platform and refers to the compatibility of the platform with 3D defined objects. In other words, the more compatible the platform is with several types of 3D objects, the more useful the platform is.

TABLE II. RESOURCES REQUIRED BY THE STUDIED FRAMEWORKS

| | | ReachIn | SOFA | CHAI3D | H3D | GiPSi | Open Haptics |
|---|---|---|---|---|---|---|---|
| Operating system | Windows | 9 | 9 | 9 | 9 | 9 | 9 |
| | Linux | 0 | 9 | 9 | 9 | 9 | 9 |
| | Mac | 0 | 9 | 9 | 9 | 0 | 9 |
| Minimum hardware requirements | HDD memory | 9 | 9 | 9 | 9 | 0 | 9 |
| | RAM memory | 9 | 1 | 9 | 9 | 0 | 3 |
| | Graphic card | 9 | 1 | 3 | 9 | 0 | 9 |
| | Processor | 3 | 1 | 3 | 9 | 0 | 9 |
| Minimum software requirements | Programming environment and compiler | 3 | 9 | 9 | 9 | 9 | 9 |
| | Graphic rendering | 3 | 9 | 3 | 3 | 3 | 9 |
| | SDK Haptic (Sensable) | 9 | 9 | 9 | 9 | 9 | 9 |
| | External libraries | 0 | 0 | 0 | 0 | 9 | 0 |

ReachIn extends the structure of VRML; from this reason, the platform may interpret this type of files. As a consequence, it may interpret the geometry of objects specified in VRML format, as IndexedFaceSet and IndexedLineSet.

On the other side, SOFA is compatible with a large number of file types that define 3D objects such as .wrl and .obj. SOFA is compatible with popular extensions like .bmp, .jpg, .png and .tiff.

CHAI3D supports both 2D, as .bmp and .tga, and 3D, as .3ds and .obj, objects. The 2D imported objects are usually used as textures for 3D objects but they may also be used to insert different graphical elements, as labels.

In order to work with 3D objects imported from 3D Studio Max or Blender, OpenHaptics provides a specialized class, named *TriMesh*, that contains lexical analyzers for .obj, .3ds, .stl and .ply files.

GiPSi platform uses only 3D objects in .obj file format and 2D resources in .tga format. Moreover, the current version of GiPSi doesn't include support for audio resources.

At last but not least, H3D is open to use both VRML and X3D standard files, as media for rapid developing and visualization of 3D virtual scenes inside a Web browser.

## D. Compatibility with haptic devices

Haptic devices are rated on a performance scale according to their characteristics.

In Table III the technical details of the haptic devices compatible with each of the studied frameworks are listed. These technical details are:

- *Haptic resolution*: The minimum distance (measured in dpi: dots per inch) between two points in the real space noticeable by the haptic device,
- *Degrees of freedom (Input - Output):* The input is the data set sent from the haptic device to the software. The output represents the force rendering process. Degrees of freedom represent the number of the translation and rotation axis.
- *Exerted force*: the maximum output force of the haptic device measured in Newton.
- *Workspace*: physical volume in which the haptic device allows movement.

Each framework is compatible with some haptic devices; for example, H3D framework is compatible with four haptic devices: Sensable, ForceDimension, Falcon, and HapticMaster.

Chai3D framework is compatible with Phantom, Delta, Falcon and Freedom6s. Additional, a class named *CustomDevice* is available and it could be used to create interaction with another type of haptic devices. Also, some frameworks can work simultaneously with two haptic devices (e.g. Chai3D and ReachIn).

In the current version, the GiPSi framework is based on the HDAPI, and it offers a programmable interface only for Phantom devices. In our review we noted that GiPSi cannot use simultaneously more than one haptic device.

For the OpenHaptic framework, the compatible devices are only the ones manufactured by Sensable: Phantom and Premium 6DOF.

## E. 3D navigation metaphors and devices

There are currently two methods for user navigation through the virtual environment navigation: pre-programmed and free navigation. While the first one needs a pre-computed path but needs no navigation device, the second gives the user the freedom to choice for viewpoint position and orientation by using an input device, usually a mouse or keyboard. This latter navigation type requires user skills in navigation and spatial orientation.

In free navigation, the user may adopt either "*view-point-in-hand*" or "*world-in-hand*" metaphors, depending on the predominant actions the user needs to execute inside the virtual environment.

Navigation is realized using desktop interaction devices (e.g. keyboard, mouse, joystick), specialized ones (e.g. spaceMouse, graphic tablet) or even haptic devices, having from 2 degrees of freedom (DOF) to 6 DOF. Depending on the navigation metaphor, the real device translation and rotation are transmitted to the viewer's viewpoint or to the objects in the scene.

Both ReachIn and H3D platforms have spaceMouse classes. The spaceMouse movements are processed before their transmission to the viewer or scene position, so that the movement maybe restricted to some specific directions, or planes.

In CHAI3D, SOFA and GiPSi, navigation is made using the mouse, based on the Glut library, or directly using the haptic device. In order to use a spaceMouse, an additional API is required.

TABLE III. TECHNICAL DESCRIPTION OF EACH STUDIED HAPTIC DEVICE

|  | Haptic resolution | Input DOF | Output DOF | Exerted force (N) | Workspace (mm) (LxIxA) |
|---|---|---|---|---|---|
| Phantom Omni | ~450 dpi | 6 | 3 | 3.3 | 160x120x70 |
| Phantom Desktop | ~1100 dpi | 6 | 3 | 7.9 | 160x120x120 |
| Phantom Premium 6DOF | >1000 dpi | 6 | 3 | 22 | 381x267x191 |
| Falcon | >400 dpi | 3 | 3 | 8.9 | 101x101x101 |
| Omega3 | >1000 dpi | 3 | 3 | 12 | 160x160x110 |
| Omega6 | >1000 dpi | 6 | 3 | 12 | 160x160x110 |
| Omega7 | >1000 dpi | 6 | 3 | 12 | 160x160x110 |
| Delta3 | >1500 dpi | 3 | 3 | 20 | 400x400x260 |
| Delta6 | >1500 dpi | 6 | 6 | 20 | 400x400x260 |
| Virtuose 6D35-45 | ~ 3100 dpi | 6 | 6 | 35 | 644x500x350 |
| Freedom6S | ~ 1500 dpi | 6 | 6 | 2.5 | 170x220x330 |
| HapticMaster | ~ 3000 dpi | 3 | 3 | 250 | 614x400x360 |
| Xitact Instrument Haptic Port | ? | 4 | 1 | 20 | 0x0x200 |

## F. Implementation language

For the development of virtual scenes each of the studied frameworks offers two levels of languages: on the low level there are the programming languages and the high level is represented by meta-languages or modeling languages. All the studied frameworks use C++ and only ReachIn and H3D allow Python scripts together with VRML in ReachIn or X3D language, in H3D. The meta-language XML is employed by SOFA, CHAI3D and GiPSi to describe the virtual scene with a graph structure.

## G. Extensibility and adaptability

The extensibility of a visuo-haptic framework consists in its capability to be continuously augmented with new functionalities. This may be obtained by the introduction of new classes or algorithms implemented by the application developer. Using dynamic typing and dynamic linking, we may change the application behavior at run-time.

According to the ISO/IEC 9126-1 [13] standard, the adaptability is defined as its capability to adapt at different media without the need of supplementary actions. Applying this concept to our project, we consider that adaptability of the platform consist in its capability to adapt to pre-defined contexts without being modified by the developer. To this end, the developer needs to make use of different levels of abstraction and parameterization.

The applications developed using ReachIn run on the basis of a scene graph structure. This entity is implemented in the C++ language using one or several classes. The developer access at implicit implementations is limited at the level of functions and class headers. During the development process new classes may be implemented based on existing ones, thus the platform is considered to be both adaptable and extensible.

CHAI3D allows the development of new C++ classes inside the API. Afterwards, these classes are available to be used in new applications.

Similarly with the previous platforms, H3D is based on the node concept, with associated fields, implemented as a C++ class. By the object-oriented derivation, the platform proves to be extensible.

GiPSi has a CORBA-based extension, named GiPSiNet [14]. That allows applications to run distributively.

In the Table IV illustrates the scores regarding these two quality properties: extensibility and adaptability.

TABLE IV. EXTENSIBILITY AND ADAPTABILITY

| | | ReachIn | SOFA | CHAI3D | H3D | GiPSi | Open Haptics |
|---|---|---|---|---|---|---|---|
| Development flexibility | Extensibility | 3 | 9 | 9 | 9 | 9 | 3 |
| | Adaptability | 9 | 3 | 1 | 3 | 3 | 3 |

*H. Real/virtual time execution*

Due to the complexity of the simulation calculus, some simulation systems need a supplementary time for the execution of each simulation time step, without the possibility to run in real time. In this direction, the virtual time [15] represents the time needed to complete execution of each simulation step. For the execution of the same simulation in real time, reducing the computational complexity is required, at the expense of the simulation fidelity.

ReachIn, CHAI3D, H3D, GiPSi and OpenHaptics platforms have been designed for developing of multisensory applications that are focused on real time execution, fast enough to be perceived by human senses as an interactive simulation.

SOFA was designed for rapid developing multisensory medical applications. This platform is characterized by strong flexibility regarding scene object definition: each object is split in several functional parts. The SOFA's scene consistency is assured by its models mappings. The models used for objects representations may be independently used. This quality offers SOFA flexibility in spite of the loss of fidelity due to the increased data processing requirement.

*I. Dynamic configuration of the scene*

This important criterion allows the evaluation of the capability of a platform/API to interact with an application structure while the application is running.

By default the platforms ReachIn, CHAI3D, H3D and GiPSi do not offer this kind of support. The only studied platform that is able to manage run-time changes in the scene graph structure is SOFA. User graphic interface gives the user the capability to interact with nodes from the scene graph structure. For example, while the application is running, the user may add or delete geometrical objects; the user may modify the values of used variables inside the scene, improving application development speed.

*J. Documentation*

While SOFA provides online documentation, ReachIn offers less support. In fact, there are four documents: two of them focus on technical support for installation, configuration and applications running; the third is the complete guide for the ReachIn classes and the last one is the programmers guide.

The CHAI3D documentation is generated using Doxygen and contains the complete class model of the platform. The documentation is completed by CHAI3D well explained examples.

OpenHaptics has a programmers guide available both for academic and non-academic use together with technical support for commercial license.

H3D API offers a very good documentation for installation and for application specific class development based on a manual, a wiki and a Doxygen generated document.

Finally, GiPSi platform is described only online [16]. Its documentation consists of several publications [8, 17] which give a good insight in the platform architecture and the motivation for its development.

*K. Availability of the API*

By the availability of a platform we understand the existence of a community and/or forum that assures an active development of the platform, or, in the case of a commercial distribution, a company that distributes it.

Currently, the company that develops the ReachIn platform stopped commercializing it a few years ago.

VI. FINAL SCORE AND DISCUSSIONS

Fig. 2 represents the frameworks/APIs evaluation together with the final scores. As we can observe, H3D API has the highest score, close followed by CHAI3D. The H3D's level score may be explained both by its popularity and by its level of support. Basis on this classification, we employed H3D framework in order to develop a visuo-haptic prototype for liver diagnostics through palpation [1].

In conclusion, the present contribution helps us in understanding and choosing the most appropriate framework/API to HapticMed project and also gives us a perspective about the lack of existing standardized APIs on the market for haptic application developers. Nevertheless, the haptic applications potential largely exceeds the potential of other existed applications because the new direction they impose in the area of simulation and training applications in key domains for society evolution (e.g. medicine, robotics, etc.).

ACKNOWLEDGMENT

Our work was supported by the ANCS grant *HapticMed - Haptic interfaces for medical applications* project, contract no 128/02.06.2010, ID/SMIS CODE 567/12271, POSCCE O.2.1.2/2009 competition. We would like to thank Aurelian Nicola, Mihai Polceanu, Petre Costin as well as the other members of the CERVA team from Ovidius University of Constanta, Romania, for their support.

| Max Relationship Value | Relative Weight | | Demanded Quality (a.k.a. "Customer Requirements" or "Whats") | Absolute importance | License | Operating system | Minimal requirements | 3D Virtual objects | 2D Virtual objects | Audio | Haptic devices | Navigation devices | Languages | Dynamic configuration of the sc | Documentation | Availability | Real/Virtual time execution | Adaptability and extensibility | ReachIn | SOFA | CHAI3D | H3D | GiPSi | Open Haptics |
|---|---|---|---|---|---|---|---|---|---|---|---|---|---|---|---|---|---|---|---|---|---|---|---|---|
| 9 | 4.000 | License | Open source licence paid | 228 | 39 | 9 | 9 | 0 | 0 | 0 | 0 | 0 | 0 | 0 | 0 | 0 | 0 | 0 | 0.00 | 0.00 | 7.40 | 9.00 | 0.00 | 0.00 |
| 9 | 2.667 | | Semi-open source license paid | 112 | 33 | 0 | 9 | 0 | 0 | 0 | 0 | 0 | 0 | 0 | 0 | 0 | 0 | 0 | 0.00 | 0.00 | 0.00 | 0.00 | 0.00 | 0.00 |
| 9 | 1.333 | | Closed source license paid | 56 | 33 | 0 | 9 | 0 | 0 | 0 | 0 | 0 | 0 | 0 | 0 | 0 | 0 | 0 | 1.25 | 0.00 | 0.00 | 0.00 | 0.0 | 0.00 |
| 9 | 6.667 | | Open source free licence | 320 | 39 | 0 | 9 | 0 | 0 | 0 | 0 | 0 | 0 | 0 | 0 | 0 | 0 | 0 | 0.00 | 7.40 | 7.40 | 9.00 | 0.00 | 0.00 |
| 9 | 5.333 | | Semi-open source free license | 224 | 39 | 0 | 3 | 0 | 0 | 0 | 0 | 0 | 0 | 0 | 0 | 0 | 0 | 0 | 0.00 | 0.00 | 0.00 | 0.00 | 7.40 | 0.00 |
| 9 | 4.000 | | Closed source free license | 96 | 21 | 0 | 3 | 0 | 0 | 0 | 0 | 0 | 0 | 0 | 0 | 0 | 0 | 0 | 0.00 | 0.00 | 0.00 | 0.00 | 0.00 | 5.50 |
| 3 | 4.000 | Minimal requir | Operation system | 28 | 0 | 7 | 0 | 0 | 0 | 0 | 0 | 0 | 0 | 0 | 0 | 0 | 0 | 0 | 3.00 | 9.00 | 9.00 | 9.00 | 6.00 | 9.00 |
| 9 | 1.333 | | Minimal hardware requirements | 24 | 0 | 0 | 18 | 0 | 0 | 0 | 0 | 0 | 0 | 0 | 0 | 0 | 0 | 0 | 7.50 | 3.00 | 6.00 | 9.00 | 0.00 | 7.50 |
| 3 | 1.333 | | Minimal software requirements | 12 | 0 | 0 | 9 | 0 | 0 | 0 | 0 | 0 | 0 | 0 | 0 | 0 | 0 | 0 | 3.00 | 4.50 | 3.00 | 3.00 | 5.25 | 4.50 |
| 9 | 6.667 | Multimodal | 3D virtual objects | 300 | 0 | 0 | 9 | 36 | 0 | 0 | 0 | 0 | 0 | 0 | 0 | 0 | 0 | 0 | 2.25 | 4.50 | 4.50 | 4.50 | 4.50 | 4.50 |
| 9 | 4.000 | | 2D virtual objects | 216 | 0 | 0 | 9 | 0 | 45 | 0 | 0 | 0 | 0 | 0 | 0 | 0 | 0 | 0 | 9.00 | 9.00 | 3.60 | 0.00 | 1.80 | 7.20 |
| 9 | 4.000 | | Audio objects | 84 | 0 | 0 | 3 | 0 | 0 | 18 | 0 | 0 | 0 | 0 | 0 | 0 | 0 | 0 | 4.50 | 0.00 | 4.50 | 0.00 | 0.00 | 0.00 |
| 9 | 6.667 | | Haptic devices | 480 | 3 | 12 | 6 | 0 | 0 | 0 | 51 | 0 | 0 | 0 | 0 | 0 | 0 | 0 | 7.62 | 2.00 | 2.92 | 6.92 | 2.07 | 2.07 |
| 9 | 5.333 | 3D navi | Eye-in-hand navigation | 112 | 0 | 0 | 0 | 0 | 0 | 0 | 0 | 21 | 0 | 0 | 0 | 0 | 0 | 0 | 9.00 | 6.00 | 6.00 | 9.00 | 6.00 | 6.00 |
| 9 | 5.333 | | World-in-hand navigation | 112 | 0 | 0 | 0 | 0 | 0 | 0 | 0 | 21 | 0 | 0 | 0 | 0 | 0 | 0 | 9.00 | 6.00 | 6.00 | 9.00 | 6.00 | 6.00 |
| 9 | 4.000 | Languages | Programming languages | 36 | 0 | 0 | 0 | 0 | 0 | 0 | 0 | 0 | 9 | 0 | 0 | 0 | 0 | 0 | 9.00 | 9.00 | 9.00 | 9.00 | 9.00 | 9.00 |
| 9 | 4.000 | | Scripting languages | 36 | 0 | 0 | 0 | 0 | 0 | 0 | 0 | 0 | 9 | 0 | 0 | 0 | 0 | 0 | 9.00 | 0.00 | 0.00 | 9.00 | 0.00 | 0.00 |
| 9 | 4.000 | | Modeling languages | 108 | 0 | 0 | 0 | 0 | 0 | 0 | 0 | 0 | 27 | 0 | 0 | 0 | 0 | 0 | 3.00 | 3.00 | 3.00 | 3.00 | 3.00 | 0.00 |
| 9 | 6.667 | | Interaction with scene graph | 80 | 0 | 0 | 0 | 0 | 0 | 0 | 0 | 0 | 0 | 12 | 0 | 0 | 0 | 0 | 0.00 | 9.00 | 0.00 | 0.00 | 0.00 | 0.00 |
| 9 | 4.000 | | Documentation | 84 | 0 | 0 | 0 | 0 | 0 | 0 | 0 | 0 | 0 | 0 | 21 | 0 | 0 | 0 | 9.00 | 3.00 | 3.00 | 9.00 | 1.00 | 9.00 |
| 9 | 4.000 | | Availability | 72 | 0 | 0 | 0 | 0 | 0 | 0 | 0 | 0 | 0 | 0 | 0 | 18 | 0 | 0 | 0.00 | 9.00 | 9.00 | 9.00 | 9.00 | 9.00 |
| 9 | 4.000 | | Simulation | 48 | 0 | 0 | 0 | 0 | 0 | 0 | 0 | 0 | 0 | 0 | 0 | 0 | 12 | 0 | 4.50 | 9.00 | 4.50 | 4.50 | 4.50 | 4.50 |
| 9 | 6.667 | | Development flexibility | 120 | 0 | 0 | 0 | 0 | 0 | 0 | 0 | 0 | 0 | 0 | 0 | 0 | 0 | 18 | 6.00 | 6.00 | 5.00 | 6.00 | 6.00 | 3.00 |
| | | | | | | | | | | | | | | | | | | Final Score | 3.917 | 3.937 | 4.050 | 5.166 | 2.771 | 2.865 |

Figure 2. House of quality of existing visuo-haptic APIs and frameworks